%% file: main.tex
\documentclass[twocolumn]{article}

\usepackage{algorithm}
\usepackage{algpseudocode}
\usepackage{amsmath}
\usepackage{amssymb}
\newcommand{\R}{\mathbb{R}}
\usepackage{authblk}
\usepackage{url}
\usepackage{derivative}
\usepackage{caption}
\usepackage{subcaption}
\usepackage{hyperref}
\usepackage{float}
\usepackage{tikz}
\usetikzlibrary{positioning, arrows.meta, shapes.geometric, 
                decorations.markings}
\usepackage{enumitem}
\usetikzlibrary{positioning, arrows.meta}
\usepackage{multirow}
\usepackage{booktabs}
\usepackage{array}
\usepackage{bm}

\usepackage[explicit]{titlesec}
\renewcommand{\thesection}{\Roman{section}}

\titleformat{\section}{\large\scshape\centering}{\thesection.\space}{0pt}{#1}[]
\titlespacing*{\section}{0pt}{0.5\baselineskip}{0pt}
\titleformat{\subsection}{\normalsize\itshape}{\Alph{subsection}.\space}{0pt}{#1}[]
\titlespacing*{\subsection}{0pt}{0.5\baselineskip}{0pt}
\titleformat{\subsubsection}{\normalsize\itshape}{\arabic{subsubsection}.\space}{0pt}{#1}[]
\titlespacing*{\subsubsection}{0pt}{0.5\baselineskip}{0pt}
\usepackage{geometry}
\newgeometry{left=2cm, right=2cm, top=2.5cm,bottom=3.5cm}
\makeatletter
\renewcommand{\fnum@figure}{Fig. \thefigure}
\renewcommand{\fnum@table}{Tab. \thetable}
\makeatother

\usepackage{nopageno}

\title{\textbf{\normalsize A practical guide to implementing zero-order-hold interplanetary trajectory legs}}
\author[(1)]{Dario Izzo}
\author[(1)]{Harry Holt}
\author[(1)]{Giacomo Acciarini}
\author[(2)]{Laurent Beauregard}
\author[(3)]{Yuri Shimane}

\affil[(1)]{Advanced Concepts Team, European Space Agency, European Space Research and Technology Centre (ESTEC), Keplerlaan 1, 2201 AZ Noordwijk, The Netherlands}
\affil[(2)]{European Space Operations Centre, Darmstadt, 64293, Germany}
\affil[]{dario.izzo@esa.int}
\affil[(3)]{Department of Mechanical and Aerospace Engineering, University of California, Irvine, CA 92617, United States of America}
\date{}  

\begin{document}
\maketitle

\begin{abstract}
\vspace{-1.3\baselineskip}
\textbf{\emph{\quad Abstract} - 
We study the practical implementation of zero-order-hold (ZOH) transcriptions for spacecraft trajectory optimisation, identifying a set of design principles that render them robust across a broad class of dynamical settings without problem-specific tuning. 
The contributions are fourfold: (i) a thorough study of the forward--backward shooting construction, denoted $\mathrm{ZOH}_\alpha$; (ii) a redundant four-dimensional throttle parameterization that eliminates the singularity of the control influence matrix along ballistic arcs; (iii) a softmax time-grid encoding that avoids ordering constraints on segment durations while preserving full differentiability; and (iv) the TOPS benchmark (Trajectory Optimisation Problems in Space), a suite of 28 problems spanning four dynamical models, two-body Cartesian, modified equinoctial elements, circular restricted three-body, and solar sailing, designed to be extended over time.
}
\end{abstract}

\section{Introduction}
We refer here to Zero-Order Hold (ZOH) transcriptions as the class of direct methods for optimal control problems in which the control input $\bm{u}(t)$ is parameterized as a piecewise-constant function and the state is computed via shooting (i.e. is resolved numerically integrating the equations of motion). Specifically, the time interval $[t_0,t_f]$ is partitioned into $N$ contiguous segments $[t_i,t_{i+1}]$, and the control is held constant at a value $\bm{u}_i$ over each segment, yielding a stepwise approximation of the continuous-time control history. 
More broadly, ZOH control parameterizations are to be viewed within the general development of direct optimal control methods, in which the functional optimisation problem is transcribed into a finite-dimensional nonlinear program (NLP) by discretizing either the control, the state, or both \cite{betts1998survey, hargraves1987direct}. 

Within low-thrust trajectory design, several distinct transcription families are often considered. 
These include direct collocation and pseudospectral methods, e.g. Hermite--Simpson, Gauss--Lobatto, Radau, and Legendre variants, which can achieve high accuracy when the optimal control is sufficiently smooth, by enforcing the dynamics at interior quadrature points and approximating the solution with higher-order polynomials \cite{topputo2014survey, fahroo2002direct, patterson2014gpops}. 
Unfortunately, in the case  of low-thrust trajectory optimisation, the smoothness hypothesis is consistently violated by the bang-bang, piecewise-constant, nature of the optimal profile dictated theoretically by Pontryagin maximum principle \cite{pontryagin1962} for minimum-fuel problem with affine control.

A ZOH parameterization, by contrast, is designed to represent piecewise-constant functions, with the switching potentially captured by the mesh topology itself. 
From this perspective, ZOH paradigms are equally capable of accurately representing optimal solutions. Their principal appeal is that they place nearly all modeling complexity in the propagation of the dynamics while keeping the control parameterization and defect construction simple. 
The Sims--Flanagan transcription \cite{sims1997preliminary,sims2006implementation}, widely used in preliminary low-thrust mission design, is a ZOH-type transcription in which each constant-thrust arc is replaced by a single impulsive $\Delta V$ applied at the midpoint of the segment, with the two half-segments propagated as Keplerian coast arcs. 
Continuous-thrust extensions of this idea, introduced later to eliminate the impulsive approximation while retaining the segmented control structure \cite{izzo2012pykep, yam2010highfidelity, yam2011lowthrust}, are among early examples of ZOH class of direct transcriptions introduced in the low-thrust trajectory optimisation domain. 

A relevant development that sits orthogonally to the classical direct-transcription taxonomy is the emergence, over the last two decades, of convexification-based solution algorithms \cite{malyuta2022convex}, most notably \emph{lossless convexification} \cite{acikmese2007convex, acikmese2012lossless} and \emph{successive convexification} (SCvx) \cite{mao2016successive, oguri2023successive}. It is important to distinguish these methods and disentangle them from transcription methods: convexification approaches are best understood as \emph{solvers}, or algorithmic wrappers, applied on top of a (possibly convex) transcription, rather than as transcription classes in their own right, in the same way that Sequential Quadratic Programming (SQP) is a solver that can be applied to the NLP arising from any transcription. In practice, within spacecraft trajectory design, the transcription over which these algorithms are most commonly deployed is precisely a ZOH-type discretization, making the present work of relevance also in that context.

Against this background, the contribution of the present work is not the introduction of a new transcription family, but the identification of a set of implementation principles that render ZOH transcriptions robust across a broad class of trajectory optimisation problems. Most previous studies, by contrast, assess performance on a small number of instances and within relatively homogeneous dynamical settings (see \cite{jiang2012practical, oshima2023regularized, englander2017automated, hofmann2023performance, dachwald2007comparison} to name a few). Consequently, the resulting methods often depend on substantial problem-specific tuning, as well as on implicit algorithmic choices whose validity is confined to the narrow application domain considered and whose manual adaptation limits scalability, automation, and reproducibility. The specific approach here presented, called ZOH$_\alpha$ is a forward backward scheme ($\alpha$ being one parameter controlling the duration of the forward and backward parts) designed to operate reliably in the absence of a problem-specific initial guess, thereby reducing the degree to which successful optimisation depends on prior intuition about the dynamics, scaling, or expected solution structure.

More specifically, we focus on four ingredients that are often treated separately in the literature but whose interaction strongly influences practical performance: (i) adaptive or variable time grids; (ii) smooth and intentionally over-parameterized representations of vector controls and boundary data; (iii) end-to-end differentiability of the full transcription, including external models such as ephemerides; and (iv) systematic benchmarking across multiple dynamical settings rather than within a single propulsion or force model. A main objective is to reduce the reliance on expert-crafted initialization, so that the transcription behaves less as a problem-dependent art requiring extensive manual intervention and more as a systematic computational methodology. The resulting goal is to position ZOH$_\alpha$ not merely as a simplistic baseline, but as a flexible and competitive transcription strategy, especially when implemented with sufficient numerical care.

\section{The fwd-bck ZOH transcription}
Let us consider a generic dynamics in the form
$$
\dot{\bm{x}} = \mathbf{f}_{\theta}(\bm{x}, \bm{u}, t), 
$$
where $\bm{x}\in\R^n$ is the state, $\bm{u}\in\mathcal{U}\in\R^m$ is the control and $\theta$ indicate possible system parameters. Let us additionally assume that we want to steer the initial system state $\bm x_0 \triangleq \bm{x}(t_0)$ to a final target state $\bm{x}_f \triangleq \bm x(t_f)$, minimizing some cost functional in the \emph{Mayer} form: $J = \phi\!\left(t_0, \bm{x}(t_0), t_f, \bm{x}(t_f)\right)$ where $\phi : \R \times \R^n \times \R \times \R^n \to \R$. This formulation entails no loss of generality: any \emph{Bolza} functional 
comprising an additional Lagrange (running cost) term 
$\int_{t_0}^{t_f} \mathcal{L}(\bm{x}, \bm{u}, t)\, dt$ 
can be reduced to the Mayer form via standard state augmentation, 
by introducing the auxiliary state $x_{n+1}$ satisfying 
$\dot{x}_{n+1} = \mathcal{L}(\bm{x}, \bm{u}, t)$, $x_{n+1}(t_0) = 0$, 
and redefining $\phi \leftarrow \phi + x_{n+1}(t_f)$.

The ZOH transcription begins by partitioning the time horizon $[t_0, t_f]$ into 
$N$ contiguous subintervals $[t_i, t_{i+1}]$, for $i = 0, \dots, N-1$. Over each 
subinterval, the control is held constant at a value $\bm{u}_i \in \mathcal{U} 
\subseteq \mathbb{R}^m$, consistently with a ZOH 
parameterization. The particular ZOH transcription considered in this work is a 
forward--backward variant, hereafter denoted by ZOH$_\alpha$, and which can be formally stated as
$$
\mathrm{ZOH}_{\alpha}:\left\{
\begin{array}{rl}
\mathrm{find:} & \bm{u}_i,\; t_j, 
\begin{array}{l}
i = 0, \dots, N-1 \\
j = 0, \dots, N 
\end{array}
\\[1mm]
\mathrm{to\ minimize:} & J(\bm{u}_i, t_j), \\[1mm]
\mathrm{subject\ to:} & \bm{mc}_\alpha(\bm{u}_i, t_j) = \bm{0}, \\
& \bm{tc}(\bm{u}_i) \le \bm{0}, \\
& \text{additional constraints.}
\end{array}
\right.
$$
Here, $\bm{mc}_\alpha$ denotes the \emph{mismatch constraints}, namely the defect 
constraints associated with the forward--backward shooting construction, while 
$\bm{tc}$ denotes the \emph{throttle constraints}, i.e., the constraints enforcing 
admissibility of the control with respect to the set $\mathcal{U}$. A visualization of the scheme is given in Figure \ref{fig:zoh_alpha}.

\subsection{The mismatch constraints}

The equality constraints $\bm{mc}_\alpha(\bm{u}_i, t_j)$ are referred to as 
\emph{mismatch constraints}. Their construction is governed by a scalar parameter 
$\alpha \in [0,1]$, hereafter called the \emph{cut}, which partitions the $N$ 
segments into
$$
N_{\mathrm{fwd}} = \lfloor \alpha N \rfloor, 
\qquad
N_{\mathrm{bck}} = N - N_{\mathrm{fwd}},
$$
that is, into $N_{\mathrm{fwd}}$ forward segments and $N_{\mathrm{bck}}$ backward 
segments, where $\lfloor \cdot \rfloor$ denotes the floor operator.

Let $\bm{x}_{\mathrm{fwd}}$ denote the state obtained by forward propagation of 
$\bm{x}_0 \triangleq \bm{x}(t_0)$ over the first $N_{\mathrm{fwd}}$ segments under 
the ZOH control sequence
$
\bm{u}_{\mathrm{fwd}} \triangleq \{\bm{u}_0, \dots, \bm{u}_{N_{\mathrm{fwd}}-1}\},
$
and let $\bm{x}_{\mathrm{bck}}$ denote the state obtained by backward propagation 
of $\bm{x}_f \triangleq \bm{x}(t_f)$ over the remaining $N_{\mathrm{bck}}$ segments 
under the reversed control sequence
$
\bm{u}_{\mathrm{bck}} \triangleq \{\bm{u}_{N-1}, \dots, \bm{u}_{N_{\mathrm{fwd}}}\}.
$
The mismatch constraints are then defined as
\begin{equation}
\label{eq:mc}
\bm{mc}_\alpha(\bm{u}_i, t_j) \triangleq 
\bm{x}_{\mathrm{bck}}(\bm{u}_{\mathrm{bck}}, t_j) -
\bm{x}_{\mathrm{fwd}}(\bm{u}_{\mathrm{fwd}}, t_j),
\end{equation}
and enforced through the condition
$
\bm{mc}_\alpha(\bm{u}_i, t_j) = \bm{0},
$
which imposes $\mathcal{C}^0$ continuity of the state trajectory at the mismatch 
point. The limiting cases $\alpha = 1$ and $\alpha = 0$ recover purely 
\emph{forward} and purely \emph{backward} shooting transcriptions, respectively.

\subsection{The throttle constraints}

The inequality constraints $\bm{tc}(\bm{u}_i)$ are referred to as 
\emph{throttle constraints}. Their precise form depends on the control 
parameterization adopted and on how the admissible control set $\mathcal{U}$ is 
represented within the transcription. Depending on the application, these 
constraints may take the form of simple nonlinear inequality constraints, equality 
constraints, or second-order cone constraints. In all cases, their role is to 
ensure that the discretized controls remain admissible with respect to the 
underlying continuous-time problem. It is convenient to think about throttles as quantities in $[0,1]$ that define the propulsion system utilization, thus decoupling them from the actual thrust available, which is not controllable and is defined very differently in nuclear electric propulsion or solar electric propulsion cases\cite{dachwald2007comparison}.

\subsection{The additional constraints}
In addition to the mismatch constraints, the transcription may include further equality and inequality constraints arising from the particular optimal control problem under consideration. Given the generality of the ZOH formulation, such constraints may encode state or path constraints, boundary conditions, and mission-specific design requirements, including, for example, fly-by periapsis constraints, limits on relative hyperbolic excess velocity, or launcher-performance constraints.

The ZOH$_\alpha$ scheme can be considered as a child of the Sims-Flanagan transcription \cite{sims2006implementation}, and bears a structural resemblance to \emph{multiple shooting}: both strategies propagate the dynamics over sub-arcs of the trajectory in order to reduce the sensitivity amplification that plagues single shooting methods. However, the ZOH$_\alpha$ formulation differs from a classical multiple shooting in two important aspects. 
First, the partition is limited to exactly two sub-arcs, one integrated forward from  $\bm{x}_0$ and one integrated backward from $\bm{x}_f$. Second, and most significantly, no additional decision variables or \emph{defect constraints} are needed since the boundary conditions of both sub-arcs are fixed a priori to the prescribed initial and final states.
This renders the transcription considerably more compact than a general multiple shooting scheme and it does not require the definition of initial guesses for intermediate states, while it retains the positive effect of decreased gradient sensitivities. The formulation has also additional advantages, mainly related to evolvability, that manifest in multi-phase problems (Low-Thrust Multiple Gravity Assist type LT-MGA) and in particular in a global optimisation setting \cite{cassioli2012global}.

\section{The Gradients}
\subsection{The mismatch constraints}
\input{figures/segments_tiks}
Computing the gradients of the mismatch constraint defined in 
Eq.~(\ref{eq:mc}) requires computing $\nabla\bm{x}_{\mathrm{fwd}}$ 
and $\nabla\bm{x}_{\mathrm{bck}}$. 
For each trajectory segment $[t_i, t_j]$, we introduce the state transition matrix (STM)
$
\bm{M}_{ji} \triangleq 
\frac{\partial \bm{x}_j}{\partial \bm{x}_i} \in \mathbb{R}^{n \times n},
$
mapping state perturbations from $t_i$ to $t_j$, and the \emph{control influence matrix}
$
\bm{T}_{ji} \triangleq 
\frac{\partial \bm{x}_j}{\partial \bm{u}_i} \in \mathbb{R}^{n \times m},
$
quantifying the sensitivity of the state at $t_j$ to the ZOH control  applied over segment $i$. Since $\bm{x} \in \mathbb{R}^n$, $\bm{M}_{ji}$ is a square $n \times n$ matrix; since $\bm{u}_i \in \mathbb{R}^m$, $\bm{T}_{ji}$ is in general a rectangular $n \times m$ matrix.
Both matrices are obtained by numerically integrating the \emph{variational equations} 
alongside the nominal trajectory. Defining
$$
\bm{A}(t) \triangleq 
\frac{\partial \mathbf{f}}{\partial \bm{x}}
\in \mathbb{R}^{n \times n},
\quad
\bm{B}(t) \triangleq 
\frac{\partial \mathbf{f}}{\partial \bm{u}}
\in \mathbb{R}^{n \times m},
$$
the STM satisfies
$$
\dot{\bm{M}}(t) = \bm{A}(t)\, \bm{M}(t), \qquad \bm{M}(t_i) = \bm{I}_n,
$$
and the control influence matrix satisfies
$$
\dot{\bm{T}}(t) = \bm{A}(t)\, \bm{T}(t) + \bm{B}(t), 
\qquad \bm{T}(t_i) = \bm{0}_{n \times m},
$$
where the control $\bm{u}_i$ is held constant over $[t_i, t_{i+1}]$ consistently with the ZOH parameterization. The values $\bm{M}_{ji} \triangleq \bm{M}(t_j)$ and $\bm{T}_{ji} \triangleq \bm{T}(t_j)$ are recovered at the end of the integration. If we then write:
$$
\bm{M}_{ji} \triangleq \bm{M}_{j(j-1)}\, \bm{M}_{(j-1)(j-2)} \cdots \bm{M}_{(i+1)i},
\quad j > i
$$ 
and $\bm{M}_{ii} \triangleq \bm{I}_n$, the chain rule takes the form (for readability, we now use the $L \triangleq N_{fwd}$ subscript indicating the second term in~\eqref{eq:mc}, i.e., the \emph{blue} triangular marker in Figure~\ref{fig:zoh_alpha}, of the mismatch constraint):
\begin{equation}
\frac{\partial \bm{x}_L}{\partial \bm{u}_i} = 
\bm{M}_{L, (i+1)}\, \bm{T}_{(i+1),i}, 
\quad i = 0, \dots, L-1,
\end{equation}
and
\begin{equation}
\frac{\partial \bm{x}_L}{\partial t_i} = 
\bm{M}_{L,i}\Bigl[
\mathbf{f}(\bm{x}_i, \bm{u}_{i-1}, t_i) - 
\mathbf{f}(\bm{x}_i, \bm{u}_{i},   t_i)
\Bigr], \quad i = 0, \dots, L,
\end{equation}
having introduced the conventions:
$$
\begin{array}{l}
\mathbf{f}(\bm{x}_0, \bm{u}_{-1}, t_0) \triangleq \bm{0} \\
\mathbf{f}(\bm{x}_{L}, \bm{u}_{L}, t_{L}) \triangleq \bm{0}
\end{array}
$$
These conventions admit a clear physical interpretation:  no arc exists before $t_0$ or after $t_{L}$ along the forward sub-trajectory, so the dynamics on those phantom sides are set to zero by definition. Similar derivations allow to write the gradient for the backward part. In some problem instances the endpoints $\bm{x}_0$ and $\bm{x}_f$ may be a function of $t_0$ and $t_f$. This situation is frequently encountered in time-free problems, and this functional relation is described by the planetary ephemerides. In these cases, the mismatch constraint gradients will have an additional term:
$$
\frac{d \bm{x}_L}{d t_0} = \frac{\partial \bm{x}_L}{\partial t_0} +  \frac{\partial \bm{x}_L}{\partial \bm{x}_0}\frac{d \bm{x}_0}{d t_0}
$$
where $\frac{d \bm{x}_0}{d t_0}$ represent the gradient of the planetary ephemerides. While recent efforts to introduce differentiability in ephemerides computations are commendable \cite{cassese2025jorbit}, they are also introducing an additional computational burden which is avoidable, in our context, by the use of a \emph{surrogate gradient}. The simplest surrogate gradient (approximation of the ephemerides sensitivity) can be obtained assuming Keplerian dynamics:
$$\frac{d\bm{x}_0}{dt_0} =
\left[
\begin{array}{c}
\frac{d\bm{r_0}}{dt_0} \\
\frac{d \bm{v_0}}{d t_0} 
\end{array}
\right] = 
\left[
\begin{array}{c}
\bm{v_0} \\
- \frac{\mu}{r_0^3}\mathbf r_0
\end{array}
\right]
$$.
In the framework of interplanetary trajectory optimisation, this has largely enough precision and the approximation introduced bears no influence on the resulting trajectory's feasibility and a negligible one on its optimality. 

\subsection{The throttle constraints}
\label{sec:throttle}
The gradients of the throttle constraints are typically straightforward to compute and introduce no additional complexity beyond the choice of thrust parameterization, which, however, deserves careful discussion. 
The throttle constraints enforce, over the $i$-th segment, the bound on the thrust magnitude and take the form $\bm{tc}(\bm{u}_i) \triangleq T(\bm{u}_i) - 1 \le 0$, where $T : \mathbb{R}^m \to \mathbb{R}$ is the throttle magnitude function. 
Their gradient reduces to $\nabla_{\bm{u}_i} T$, whose form depends critically on how the throttle vector is parameterized by $\bm{u}_i$.
A natural choice is to choose $\bm{u}_i$ directly as the Cartesian components of the throttle vector, so that $T = \|\bm{u}_i\|_2$. 
This introduces a well-known singularity: $\nabla_{\bm{u}_i} T$ diverges as $T \to 0$, rendering ballistic arcs a singular locus of the parameterization. 
Replacing the constraint with $T^2 \le 1$ does not resolve the issue: the throttle magnitude $T$ appears directly in the equations of motion through the mass-flow equation (e.g. $\dot m = -c T$), so the control influence matrix 
$\bm{B}(t) = \frac{\partial \mathbf{f}}{\partial \bm{u}}$ retains a term proportional to $\nabla_{\bm{u}} T$, which remains singular at $T = 0$ regardless of the constraint formulation.
The singularity is therefore structural and must be addressed at the level of the parameterization itself.
To this end, we adopt a redundant four-dimensional parameterization. Over 
segment $i$, the throttle vector is written as
$$
\bm{T}^{(i)} \triangleq \tau^{(i)}\,\bm{d}^{(i)}, 
\qquad \bm{d}^{(i)} \in \mathbb{R}^3,
$$
yielding the augmented control $\bm{u}_i \triangleq (\tau^{(i)}, \bm{d}^{(i)}) \in \mathbb{R}^4$. 
Let us then define the admissible control set as:
$$
\mathcal{U} \triangleq \bigl\{(\tau, \bm{d}) : \tau \in [0, 1],\;
\|\bm{d}\|_2 \le 1\bigr\}.
$$
where the inequality constraint on the norm of $\bm{d}$ is to be noted.
Under these choices, the throttle constraint is in principle $\tau^{(i)}\|\bm{d}^{(i)}\|_2 \le 1$, but a simpler form is possible.
Since $\mathcal{U}$ already enforces $\|\bm{d}^{(i)}\|_2 \le 1$, we have
\begin{equation}
    \tau^{(i)}\|\bm{d}^{(i)}\|_2 \le \tau^{(i)} \cdot 1 = \tau^{(i)},
\end{equation}
so the condition $\tau^{(i)} \le 1$ implies the original constraint for all admissible $\bm{d}^{(i)}$, making it a valid sufficient condition; one that, as we shall prove, is not actually restrictive. Combined with the natural lower bound $\tau^{(i)} \ge 0$ (non-negative throttle), the throttle is confined to the unit interval $\tau^{(i)} \in [0,1]$ which is trivial to deal with.
Under this parameterization, we write, for each segment, the mass-flow equation as $\dot{m} = -c\,\tau^{(i)}$, where we intentionally leave the term $\|\bm{d}\|_2$ out.
This violates, in principle, the dynamics since it allows for a mass flow larger than the one dictated by the actual throttle norm, an effect modeling a mass dumping strategy, (this happens when $\|\bm{d}\|_2<1$), but results in a smooth control influence matrix $\bm{B}(t)$ even along ballistic arcs ($\tau^{(i)} = 0$).
In reality, this formulation is fully consistent with the original dynamic.
By the Pontryagin Maximum Principle~\cite{pontryagin1962}, the Hamiltonian of our mass-optimal low-thrust problem contains the term $\bm{d} \cdot \bm{\lambda}_v$, where $\bm{\lambda}_v$ is the costate associated with velocity. Maximizing over $\mathcal{U}$ requires $\|\bm{d}\|_2 = 1$, hence forbidding mass dumping as a strategy for optimal solutions.
In other words, the feasible sets of the original problem and our simplified form coincide at any optimal solution, and any first-order critical point of the simplified NLP is also a critical point of the original.
In this context, throttle constraints may therefore equivalently be defined with the inequality $\tau^{(i)}\le 1$ or the equality  $\tau^{(i)} = 1 $; both formulations are tested in the numerical experiments reported as they test differently solvers mechanics, such as active set strategies and barrier functions, and thus have a potential significant effect on their efficiency.

\section{The time grid}

The simplest time grid is uniform, defined by a single parameter 
$t_{\mathrm{tof}} \triangleq t_f - t_0$:
$$
t_i = t_0 + i\,\frac{t_{\mathrm{tof}}}{N}, \qquad i = 0, \dots, N.
$$
While robust, a uniform grid forces switching points to fall at 
arbitrary interior locations within segments, potentially degrading 
the representation of bang-bang control structures. A variable time 
grid is therefore preferable.
Two approaches are discussed. The first introduces the grid nodes 
$t_i$ directly as decision variables, subject to the ordering 
constraints $t_i \le t_{i+1}$. This yields a smooth transcription 
with variable-length segments at the cost of $N+1$ additional 
variables and $N$ additional affine inequality constraints.
The second avoids the inequality constraints by parameterizing the 
segment durations via the softmax transformation. Introducing an 
unconstrained vector $\bm{w} \in \mathbb{R}^N$, the normalized 
durations are defined as
$$
\Delta t_i \triangleq \frac{e^{w_i}}{\sum_{j=0}^{N-1} e^{w_j}},
\qquad i = 0, \dots, N-1,
$$
so that $\sum_i \Delta t_i = 1$ is satisfied identically for any 
$\bm{w}$, and the physical durations are recovered as 
$h_i = t_{\mathrm{tof}} \cdot \Delta t_i$. This map has three 
desirable properties: it is a smooth bijection from $\mathbb{R}^N$ 
onto the interior of the probability simplex, guaranteeing strict 
positivity $h_i > 0$; its Jacobian is full rank everywhere, 
preserving gradient information; and it requires no additional 
affine constraints in the NLP, albeit at the higher nonlinearity introduced by the softmax transformation. We test this softmax representation as well as the uniform time grid transcription.

\section{Benchmark problem definitions}
\begin{figure*}[tb]
    \centering
    \includegraphics[width=0.85\linewidth]{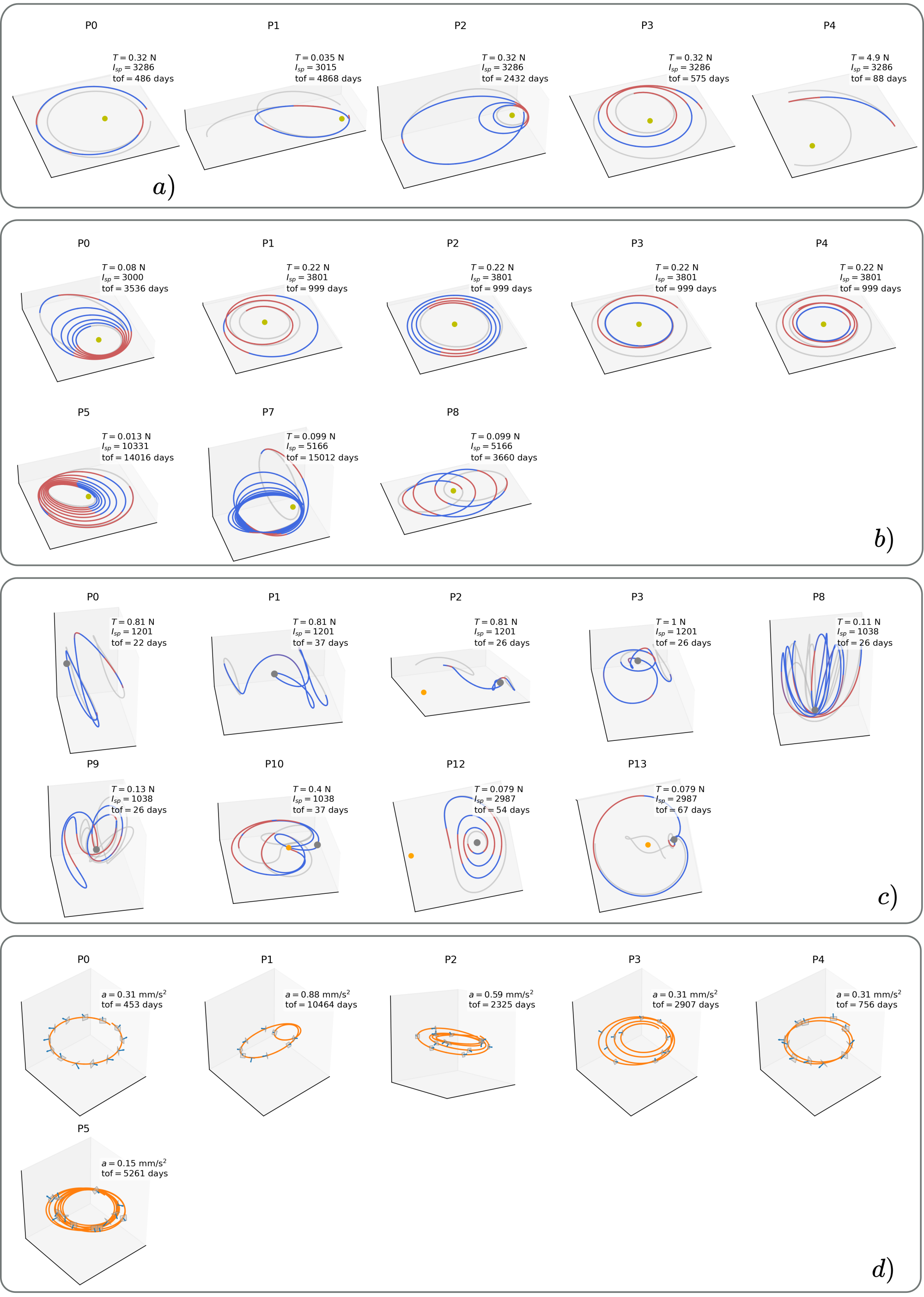}
    \caption{Summary of all problems in the TOPS benchmark. a) \emph{twobody} the basic Cartesian spacecraft dynamics, b) \emph{mee} same, but using Mean Equinoctial Elements, c) \emph{cr3bp} the circular restricted three body problem and d) \emph{ss} the solar sailing dynamics.}
    \label{fig:benchmark}
\end{figure*}
We select four different, well-studied spacecraft dynamics: \emph{twobody}, \emph{mee}, \emph{cr3bp}, and \emph{ss}. They represent the controlled motion of a spacecraft, respectively, in Cartesian elements, in modified equinoctial elements, in the circular restricted three-body problem, and equipped with a solar sail. For each dynamics we define multiple trajectory problems capturing different levels of difficulties and challenges for a total of 28 problems, which we plan to further grow in future works and that we refer to as the TOPS benchmark (Trajectory Optimisation Problems in Space). The problems cover multiple revolutions, inclination changes, as well as high eccentricities, to provide a complete set of interesting and challenging instances. The exact problem definitions, as well as all details on the dynamics used, are available in the open source project \emph{zoh}, which is publicly released\footnote{\url{https://gitlab.com/EuropeanSpaceAgency/zero-order-hold}} and that contains all the necessary code and information to reproduce our results. Some of the problems in our TOPS benchmark are inspired by previous publications by various authors \cite{taheri2016enhanced, jiang2012practical, oshima2023regularized}, most of which have been introduced to form a complete set of benchmarks. A visual summary of all the problem instances considered, including the best found trajectory, is provided in Figure \ref{fig:benchmark}. Note that in the figure, units are rescaled to solar system equivalents so that a GTO-GEO transfer (Earth system) would be visualized as the equivalent heliocentric transfer.

\section{Experimental Setup}

\subsection{Protocol}
\label{sec:protocol}
For each of the 28 problems in the TOPS benchmark, we build a $\mathrm{ZOH}_\alpha$ transcription with $N \in \{5, 10, 20, 40\}$ segments (extended to $N \in \{60, 80, 100\}$ for the more demanding instances), $\alpha \in \{0, 0.5, 1\}$, both uniform and softmax time grids, and throttle constraints enforced as either equalities or inequalities as described in Section~\ref{sec:throttle}. In the case of the \emph{ss} (solar sail) problems only we set $N \in \{9,11,13,15,17,19,21,23,25,27,29,31\}$, and we do not have throttle constraints as propulsion is created by the solar sail. To solve all dynamics and its variational version (for the gradient computations) we use our own modern Taylor integrator (Heyoka \cite{biscani2021revisiting}) setting tolerances to $10^{-16}$ for the dynamics and $10^{-8}$ for the variational dynamics.

For each configuration, 128 random initial guesses are drawn independently and uniformly from the admissible control set $\mathcal{U}$.
Each instance in \emph{twobody}, \emph{mee} and \emph{ss} is first scaled to non dimensional units so that $\mu=1$ and $|\bm{r}_0|=1$. This is of great importance to the solver performances. Then, the instance is solved to termination with both IPOPT \cite{ipopt} and SNOPT \cite{snopt}. No solver-specific tuning is performed: both solvers are run with default configurations, with the sole exception of the stopping tolerances, which are set as follows. For SNOPT, both feasibility and optimality tolerances are set to  $10^{-10}$, while the maximum number of major iterations is set to $2000$. For IPOPT, the tolerance is set to $10^{-10}$ while the maximum number of iterations to $2000$.
A run is declared \emph{successful} if the returned solution is feasible at termination, i.e.\ if the infinity norm of the constraint violation does not exceed~$\delta_{\mathrm{feas}} = 10^{-7}$ (this relaxes the $10^{-10}$ requirement and avoids declaring as failed runs due to uncontrolled inner implementation details of the solvers).

\subsection{Performance Metrics}

During each run $j$ we record (i)~the final spacecraft mass~$J$ (the objective), (ii)~the number of objective-and-constraint  evaluations~$n_f^{(j)}$, and (iii)~the number of gradient evaluations~$n_g^{(j)}$.
These counts are combined into a single hardware-normalised cost via the \emph{expected run time}~(ERT), defined following the standard practice in benchmarking stochastic and multistart optimisers~\cite{mersmann2015analyzing} as
\begin{equation}
  \mathrm{ERT}
  \;\triangleq\;
  \frac{1}{N_{\mathrm{succ}}}
  \sum_{i=1}^{N_{\mathrm{run}}}
  \bigl(n_f^{(i)} + \eta\,n_g^{(i)}\bigr)\,c ,
  \label{eq:ert}
\end{equation}
where $N_{\mathrm{succ}}$ is the number of successful runs, $N_{\mathrm{run}} = 128$ is the total number of runs, $c$~[s] is the wall-clock cost of a single objective-and-constraint evaluation on the experimental platform, and $\eta c$~[s] is the corresponding cost its gradient evaluation.
Both $c$ and $\eta$ are estimated once by timing representative evaluations prior to the benchmark and are then held constant.
Using a metric based on evaluation counts rather than raw wall-clock time makes it portable and robust to transient load variations; the conversion back to seconds is exact through the factors $c$ and $\eta c$.
Crucially, unsuccessful runs are \emph{not discarded}: they contribute their full evaluation costs to the numerator, so $\mathrm{ERT}$ grows monotonically as the success rate decreases, correctly penalising setups that fail frequently.

A second scalar of interest is the \emph{expected objective} among successful runs,
\begin{equation}
  \mathbb{E}[J]
  \;=\;
  \frac{1}{N_{\mathrm{succ}}}
  \sum_{i \in \mathcal{S}} J_i ,
  \label{eq:meanobj}
\end{equation}
where $\mathcal{S}$ denotes the index set of successful runs. For most problems in $\mathcal{B}$, the objective is the final spacecraft mass ratio; higher values therefore correspond to higher-quality solutions.

\subsection{Benchmarking Methodology}

Solver and transcription performance is characterised by the \emph{non-dominated front} in the $\bigl(\mathrm{ERT},\,\mathbb{E}[J]\bigr)$ plane, computed separately for each problem $\mathcal{P}$ in TOPS and aggregated across the suite.
A configuration $A$ \emph{dominates} configuration $B$ on a given problem if it achieves a lower ERT and a higher (or equal) $\mathbb{E}[J]$, or a strictly higher $\mathbb{E}[J]$ at equal ERT.
The resulting Pareto front simultaneously exposes the \emph{speed} and the \emph{solution quality} of each solver--transcription pair without collapsing either dimension into a scalar threshold.

This approach departs from the classical performance-profile framework of Dolan and
Mor\'{e}~\cite{dolan2002benchmarking}, which summarises solver performance through a one-dimensional cumulative distribution function of a performance ratio evaluated at a fixed quality threshold.
That construction is well-suited to problems possessing a unique, well-identified optimum, where a binary success criterion is unambiguous.
Our setting differs in two fundamental respects.
First, the trajectory problems in $\mathcal{B}$ are non-convex NLP that admit multiple local minima of some interest, so any fixed quality threshold is inherently arbitrary and its choice would materially affect the resulting ranking.
Second, and more importantly, we are explicitly interested in solver behaviour across the entire quality--cost spectrum: a configuration that is moderately suboptimal but substantially faster than the best-known solution is genuinely useful for mission-design exploration, yet it would be classified as a failure under any fixed threshold that lies below its objective value.
The Pareto-front formulation retains this information by construction, and it subsumes the Dolan--Mor\'{e} profile as a one-dimensional projection at any chosen accuracy level, while preserving the full tradeoff across all levels simultaneously.

\section{Results}
\begin{figure*}[tb]
    \centering
    \includegraphics[width=0.95\linewidth]{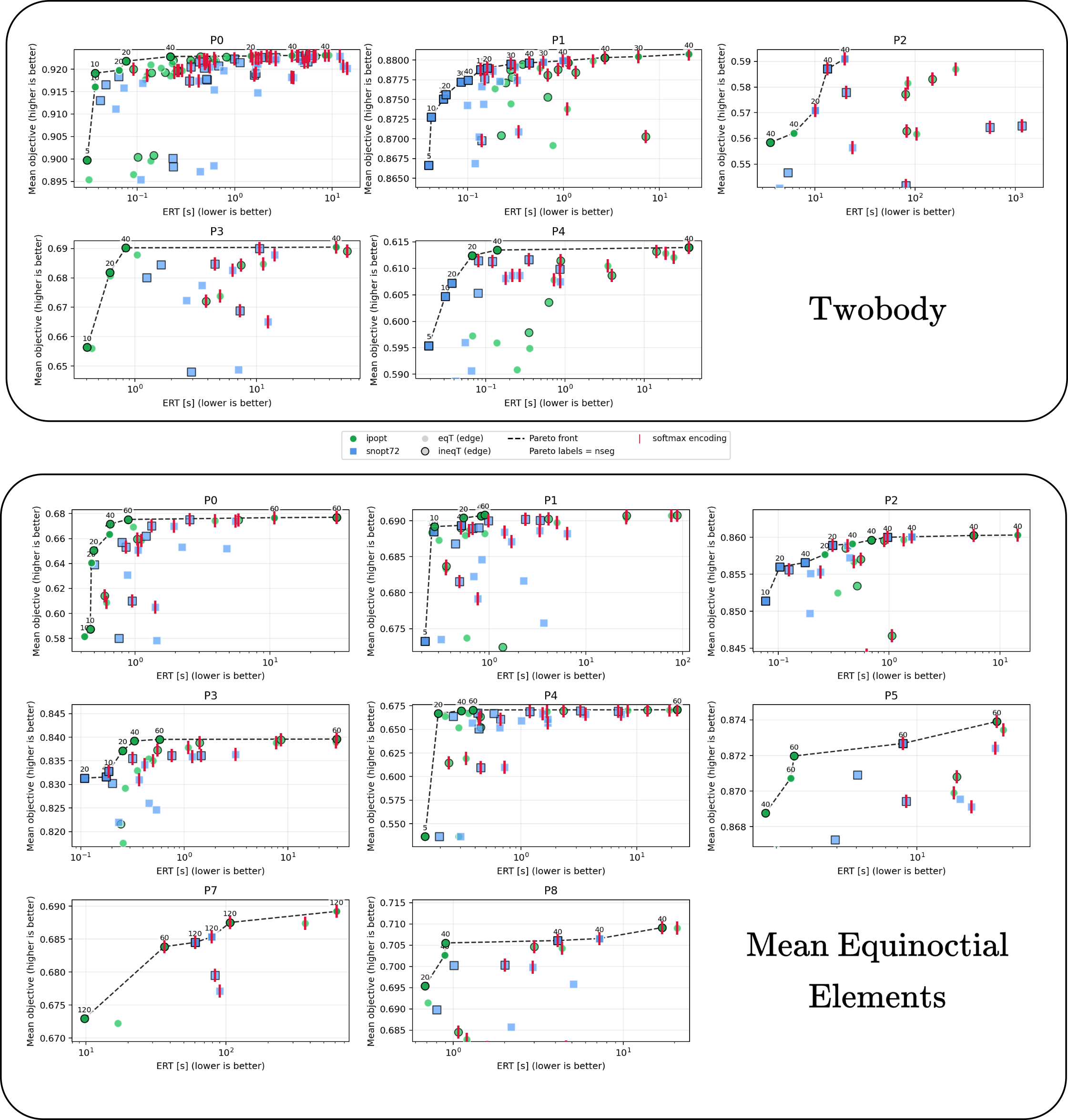}
    \caption{The non-dominated fronts in the mean objective, ERT [s] metrics for all the TOPS problems in the \emph{twobody} and \emph{mee} dynamics.}
    \label{fig:tb_mee}
\end{figure*}
\begin{figure*}[tb]
    \centering
    \includegraphics[width=0.95\linewidth]{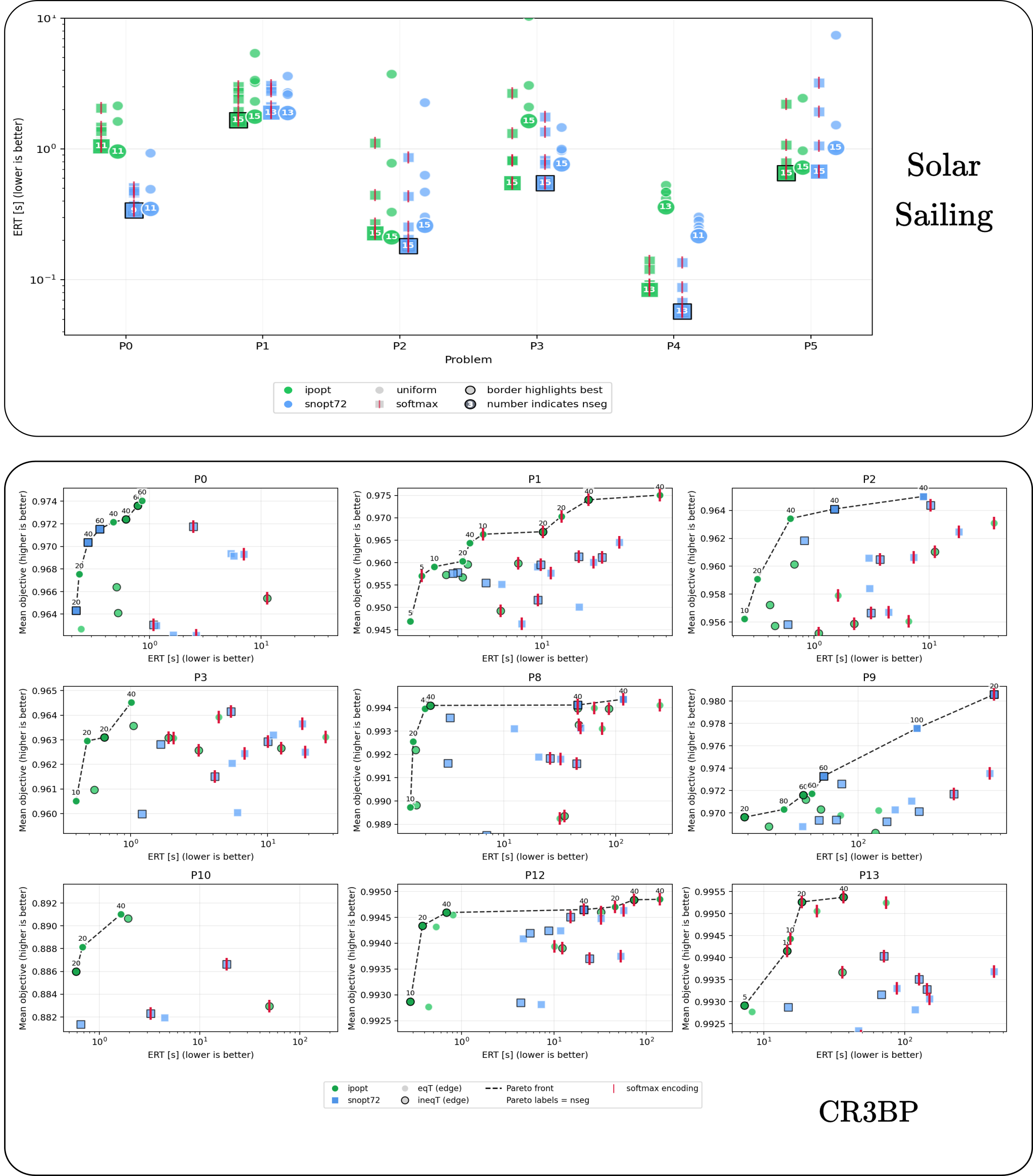}
    \caption{top: Performance plots for all the TOPS problems in the \emph{ss} dynamics (constraint satistaction, only metric is ERT).
    bottom: The non-dominated front in the mean objective, ERT [s] metrics for all the TOPS problems in the \emph{cr3bp}.}
    \label{fig:tb_cr3bp}
\end{figure*}

The experimental protocol described in Section~\ref{sec:protocol} 
produced in excess of 300{,}000 solver runs, visualized in figures \ref{fig:tb_mee}-\ref{fig:tb_cr3bp} and yielding a number of interesting findings; we report here the most significant ones.
\subsection{fwd vs bck vs fwd-bck}
In all non-dominated fronts computed for each benchmark problem, transcriptions with $\alpha = 0.5$ (forward--backward) consistently dominated those with $\alpha = 1$ (purely forward) and $\alpha = 0$ (purely backward). 
This is not visualized in figures \ref{fig:tb_mee}-\ref{fig:tb_cr3bp} as we chose to not visualize differently runs with different $\alpha$, to avoid cluttering, but this result is robust across problems and dynamics,
and strongly suggests fixing $\alpha = 0.5$ and removing it from the set of hyperparameters of the transcription altogether. 
The performance advantage is attributable to two concurring factors: the reduced gradient sensitivity inherent to the forward--backward shooting scheme, and the more informative use of random initial guesses, which 
in the forward--backward case simultaneously exploit both boundary conditions $\bm{x}_0$ and $\bm{x}_f$ rather than one alone.

\subsection{IPOPT vs SNOPT}
A comparison of IPOPT and SNOPT reveals a balanced outcome across the benchmark suite, with the notable exception of solar sail problems where SNOPT seems to have a consistent advantage. 
This is noteworthy given that SNOPT is the de facto NLP solver in aerospace trajectory optimisation, underpinning widely-used tools such as GPOPS-II and DIDO~\cite{betts1998survey, fahroo2002direct, patterson2014gpops, snopt, rao2009survey}, while IPOPT is often overlooked as a possible tool.
We emphasise, however, that this result is contingent on two features of our experimental setup: full differentiability of the transcription and the use of scaled problem instances. 
In preliminary experiments where either condition was not met, SNOPT exhibited comparatively stronger performance, suggesting that SNOPT's internals are more tolerant of poor scaling and non-smooth dynamics than IPOPT's.
Several structural factors are likely responsible for IPOPT success. 
First, IPOPT's logarithmic barrier method handles all inequality constraints simultaneously and does not require an initial active-set guess, making it inherently more robust to cold starts from random initial guesses far from feasibility. 
Additionally, the Wächter--Biegler filter line search with second-order corrections is specifically designed to avoid the Maratos effect near the constraint manifold~\cite{ipopt}, making it well-suited to the highly nonlinear shooting mismatch constraints where linearization quality degrades away from the solution.
We note that neither solver was tuned and both were run with default settings, which may favour a solver in this particular problem class; the relative performance of problem-specific SNOPT and IPOPT configurations remains an open question.

\subsection{Uniform vs variable time grid}
As expected, variable time-grid transcriptions achieve generally higher accuracies at the cost of extra computation, consistently occupying the upper-right region of the Pareto front.
This suggests their natural use as a \emph{finisher}: a warm-started refinement applied to initial solutions obtained with a uniform time grid.

An unexpected and particularly interesting exception emerges from the solar sailing problems, for which constraint satisfaction is the sole success criterion and ERT the only performance metric. 
In those instances, the softmax time-grid encoding consistently outperforms the uniform one, suggesting a fundamentally different interaction between the transcription and the underlying dynamics.

Solar sailing dynamics differs substantially from the low-thrust, mass-varying case: there is no mass equation, and none of the associated numerical characteristics, and the controls are the sail 
clock and cone angles, which introduce their own gradient degeneracies near ballistic solutions. It is therefore not surprising that solvers and transcriptions behave differently in 
this setting. 
A plausible explanation for the superiority of the non-uniform time grid is that the softmax parameterization grants the transcription the freedom to contract segments with detrimental sail orientation. Unlike for other problems where the magnitude and individual control directions are free to be chosen independently, for a solar sail the control directions are coupled: an advantageous control in one direction might be coupled to a detrimental one in another. The flexibility to reduced or relocate these harmful configurations can reduces their adverse effect on convergence.

\subsection{Equalities vs Inequalities for Throttle Constraints}
Transcriptions using the inequality form of the throttle constraint consistently dominate their equality-based counterparts in the non-dominated fronts, with one notable exception: problem instances in which the optimal trajectory is predominantly ballistic and optimal powered arcs are shorter than the corresponding segment duration.
In such cases, the equality formulation performs comparably or better. Two concurring factors explain this: first, equality constraints prevent the throttle direction vector from losing unit-norm regularity near singular arcs, preserving gradient informativeness throughout the optimisation;
second, in these problem instances the optimal $\tau$ is interior to $[0,1]$ along powered arcs, meaning that neither the active-set logic of SQP nor the logarithmic barrier gains its typical structural advantage, levelling the playing field between the two constraint formulations.
We therefore recommend the inequality form as the default for $\mathrm{ZOH}_\alpha$ transcriptions, reserving equality constraints for problems where prior knowledge suggests that the engine delivers only a small number of well-placed impulsive-like arcs.

\bibliographystyle{ISSFD_v01}
\bibliography{bibliography}

\end{document}

%% file: figures/segments_tiks.tex

\begin{figure*}[t]
\centering
\begin{tikzpicture}[
    >=Stealth,
    scale=1.3,
    dot/.style={circle, fill=black, inner sep=0pt, minimum size=5pt},
    thrustfwd/.style={-{Stealth[length=4pt,width=3pt]},
                      very thick, green!60!black},
    thrustbck/.style={-{Stealth[length=4pt,width=3pt]},
                      very thick, orange!80!black},
    ulabel/.style={font=\footnotesize, fill=white,
                   draw=none, inner sep=1.5pt},
]

\coordinate (x0)   at (0.0, 0.00);
\coordinate (n1)   at (1.5, 0.90); 
\coordinate (n2)   at (4.8, 0.60); 
\coordinate (xfwd) at (6.3, 1.10); 
\coordinate (xbck) at (6.3, 1.80); 
\coordinate (n4)   at (7.4, 2.30); 
\coordinate (xf)   at (9.5, 1.70); 

\tikzset{
  thrust marks fwd/.style={
    postaction={decorate, decoration={markings,
      mark=at position 0.12 with {
          \draw[thrustfwd,rotate={#1-\pgfdecoratedangle}](0,0)--(0.84,0);},
      mark=at position 0.30 with {
          \draw[thrustfwd,rotate={#1-\pgfdecoratedangle}](0,0)--(0.84,0);},
      mark=at position 0.50 with {
          \draw[thrustfwd,rotate={#1-\pgfdecoratedangle}](0,0)--(0.84,0);},
      mark=at position 0.70 with {
          \draw[thrustfwd,rotate={#1-\pgfdecoratedangle}](0,0)--(0.84,0);},
      mark=at position 0.88 with {
          \draw[thrustfwd,rotate={#1-\pgfdecoratedangle}](0,0)--(0.84,0);},
    }}
  },
  thrust marks bck/.style={
    postaction={decorate, decoration={markings,
      mark=at position 0.12 with {
          \draw[thrustbck,rotate={#1-\pgfdecoratedangle}](0,0)--(0.84,0);},
      mark=at position 0.30 with {
          \draw[thrustbck,rotate={#1-\pgfdecoratedangle}](0,0)--(0.84,0);},
      mark=at position 0.50 with {
          \draw[thrustbck,rotate={#1-\pgfdecoratedangle}](0,0)--(0.84,0);},
      mark=at position 0.70 with {
          \draw[thrustbck,rotate={#1-\pgfdecoratedangle}](0,0)--(0.84,0);},
      mark=at position 0.88 with {
          \draw[thrustbck,rotate={#1-\pgfdecoratedangle}](0,0)--(0.84,0);},
    }}
  },
}


\draw[blue!80!black, very thick,
    thrust marks fwd=60,
    postaction={decorate, decoration={markings,
        mark=at position 0.60 with {\arrow{Stealth}}}}
] (x0) .. controls (0.6, 0.9) and (1.1, 0.5) .. (n1);

\draw[blue!80!black, very thick,
    thrust marks fwd=20,
    postaction={decorate, decoration={markings,
        mark=at position 0.60 with {\arrow{Stealth}}}}
] (n1) .. controls (2.6, 1.4) and (3.9, 0.2) .. (n2);

\draw[blue!80!black, very thick,
    thrust marks fwd=-70,
    postaction={decorate, decoration={markings,
        mark=at position 0.60 with {\arrow{Stealth}}}}
] (n2) .. controls (5.4, 0.8) and (5.9, 0.5) .. (xfwd);


\draw[red!70!black, very thick,
    thrust marks bck=250,
    postaction={decorate, decoration={markings,
        mark=at position 0.60 with {\arrow{Stealth}}}}
] (xf) .. controls (9.2, 2.4) and (8.2, 2.0) .. (n4);

\draw[red!70!black, very thick,
    thrust marks bck=80,
    postaction={decorate, decoration={markings,
        mark=at position 0.60 with {\arrow{Stealth}}}}
] (n4) .. controls (7.2, 2.5) and (6.8, 1.9) .. (xbck);

\foreach \p in {x0, n1, n2, n4, xf} {
    \node[dot] at (\p) {};
}

\node[regular polygon, regular polygon sides=3,
      draw=blue!80!black, fill=blue!20, inner sep=2pt] at (xfwd) {};
\node[regular polygon, regular polygon sides=3,
      draw=red!80!black, fill=red!20,
      inner sep=2pt, rotate=180] at (xbck) {};

\draw[<->, densely dashed, gray!60!black, thin]
    ([yshift=5pt] xfwd) -- ([yshift=-5pt] xbck)
    node[midway, right=4pt, font=\footnotesize, text=gray!80!black]
    {$\bm{mc}_\alpha$};

\node[ulabel, text=green!60!black]  at (0.85, 1.35) {$\bm{u}_0$};
\node[ulabel, text=green!60!black]  at (3.40, 1.1) {$\bm{u}_1$};
\node[ulabel, text=green!60!black]  at (5.85, -0.12) {$\bm{u}_2$};
\node[ulabel, text=orange!80!black] at (8.45, 1.4) {$\bm{u}_4$};
\node[ulabel, text=orange!80!black] at (6.75, 2.65) {$\bm{u}_3$};

\node[below left=2pt,  font=\footnotesize] at (x0)   {$\bm{x}_0,\, t_0$};
\node[below=4pt,       font=\footnotesize] at (n1)   {$\bm{x}_1,\, t_1$};
\node[below=5pt,       font=\footnotesize] at (n2)   {$\bm{x}_2,\, t_2$};
\node[above right=4pt, font=\footnotesize] at (n4)   {$\bm{x}_4,\, t_4$};
\node[below right=2pt, font=\footnotesize] at (xf)   {$\bm{x}_f,\, t_f$};
\node[right=2pt,       font=\footnotesize] at (xfwd) {$\bm{x}_{\mathrm{fwd}},\, t_3$};
\node[left=2pt,        font=\footnotesize] at (xbck) {$\bm{x}_{\mathrm{bck}},\, t_3$};

\node[text=blue!80!black, font=\small] at (3.2, -0.35)
    {forward ($N_{\mathrm{fwd}} = 3$)};
\node[text=red!80!black,  font=\small] at (7.8,  3.05)
    {backward ($N_{\mathrm{bck}} = 2$)};

\end{tikzpicture}
\caption{Illustration of the $\mathrm{ZOH}_\alpha$ forward--backward
transcription with $N=5$, $N_{\mathrm{fwd}}=3$, $N_{\mathrm{bck}}=2$.
Blue: forward arc propagated from $\bm{x}_0$; red: backward arc
propagated from $\bm{x}_f$. Green and orange arrows indicate the constant
ZOH thrust $\bm{u}_i$ per segment. Triangular markers represent
$\bm{x}_{\mathrm{fwd}}$ (blue, upward) and $\bm{x}_{\mathrm{bck}}$
(red, downward) at the shared junction time $t_3$; the dashed gap
is the mismatch $\bm{mc}_\alpha$.}
\label{fig:zoh_alpha}
\end{figure*}